# Auditory verbal learning disabilities in patients with mild cognitive impairment and mild Alzheimer's disease: A clinical study

Mahmoud Kargar[a]* and, Alireza Mohammadi[b]

[a] Department of Speech Therapy, School of Rehabilitation, Tehran University of Medical Sciences, Tehran, Iran.

[b] Neuroscience Research Center, Baqiyatallah University of Medical Sciences, Tehran, Iran

## Abstract

Learning and memory impairments are common characteristics of individuals with mild cognitive impairment (MCI) and mild Alzheimer's disease (miAD). Early diagnosis of MCI is necessary to prevent recurrence of the disease and developing of miAD. For this purpose, we investigated the components of the Rey Auditory-Verbal Learning Test (RAVLT) to explore the auditory-verbal learning (AVL) disabilities in these patients. The AVL of 20 patients with miAD and 30 patients with MCI were compared with 30 cognitively normal controls (CN) using the RAVLT. General cognitive performance assessment was carried out based on the Mini-Mental State Examination (MMSE) score. Finally, Pearson's correlation coefficients were used to evaluate the correlation between the MMSE scores and immediate and delayed recalls, verbal learning and forgetting, and memory recognition. We found that both miAD and MCI subjects were significantly impaired in all components of the RAVLT. Compared to the MCI subjects, miAD patients performed worse on all components of the test. The MCI subjects had significantly lower scores than the CN group. The AVL analysis showed that there were significant differences between the CN and other groups, but the difference between MCI and miAD subjects was not significant. However, there was no difference among the groups in their verbal forgetting scores. It can be concluded that both patients with miAD and MCI were impaired in AVL and our findings confirm that the RAVLT can take a part in the prediction of probable miAD and early evaluation of individuals with subjective memory complaints.

# 1. Introduction

Memory is defined as the ability to maintain and use information for achieving appropriate goals (Fuster, 1999). Appropriate performance of memory depends on the health of several regions in the brain, including areas that are very sensitive to injury or disease. Learning and memory impairments are common complaints of individuals suffering from neurological disorders which impact daily activities and functional capabilities (Lezak, 2012; Malec & Thompson, 1994; Schoenberg et al., 2006). Memory impairment is considered as the most important reason for seeking neuropsychological evaluation by patients (Lezak, 2012). In addition, many psychiatric and neurological diseases cause impairments in the process of memory (Bozikas et al., 2006; Mitrushina et al., 1991; Schoenberg et al., 2006). MCI is a clinical state between normal aging and dementia which develops into miAD in many cases (Mohammadi et al., 2018). The global prevalence of AD has been reported to be around 44 million in 2016, and the number of patients is set to double every 20 years (Javadpour & Mohammadi, 2016). Neuropsychological, neuropathological, and neuroimaging studies have also shown the transient nature of MCI. MCI patients with objective memory deficits are approximately four times more likely to develop progressive AD than those with a normal cognitive status (Petersen et al., 1999). Although MCI is the initial phase of cognitive impairment that slightly affects daily functions, these individuals are still able to live independently (Golob et al., 2002; Papaliagkas et al., 2011). The prevalence of MCI in the elderly population (50 to 95 years) is 5.8 to 18.5 percent. Patients with MCI have a higher proportion of verbal interaction (nearly 50 percent) compared to patients with dementia.

Hence, MCI, as a transient state between normal aging and dementia, is considered to be an important target for the treatment of dementia (Ally et al., 2009; Villeneuve et al., 2009). It has been shown that these individuals have an increased risk of the progressive AD (1−25 percent per year) (Dawe et al., 1992). Early detection and accurate differential diagnosis of AD are necessary to develop novel treatments and improve palliative therapies. Early neuropsychological, biological, and neurochemical markers are powerful tools for achieving this goal (Estévez-González et al., 2003). The RAVLT is a useful neuropsychological test for measuring the ability of encoding, consolidation, and storage of verbal information (Bigler et al., 1989; Rey, 1941; Schmidt, 1996; Taylor, 1959). Also, the RAVLT is known as a sensitive test for verbal learning and memory evaluation (Bigler et al., 1989; Drebing et al., 1994; Ivnik et al., 1994; Mungas, 1983; Powell et al., 1991; Tuokko et al., 1995). It has been found that its performance is affected by age, education, intelligence, and gender. Whereas, it is unaffected by gender (Schmidt, 1996). The impacts of education, intelligence, and sex on the performance of the RAVLT interfere with each other, but it is generally accepted that education and intelligence affect the performance on the RAVLT and women perform slightly better than men (Schmidt, 1996; Uchiyama et al., 1995).

A new interesting trend in these types of clinical studies, is using EEG signal to better analyse the EEG and possible anomalies and also modelling the regional localization and connectivity as well (Mohamadi et al., 2017; Mohamadi et al., 2019). Authors of reference (Mohamadi et al. 2017) developed a very comprehensive framework allowing for better analysing EEG signal in different problem setting. In his framework, they suggest to model both linear and nonlinear characteristics of EEG signal for diagnostic purposes such as epileptic seizures. However, they argue that the framework is extendable to multichannel EEG analysis for accurate localization of involved regions. In this regard, also this work could be used as a complementary analysis in our clinical studies.

The RAVLT is sensitive to verbal memory deficits in various diseases, including AD, Parkinson's disease (PD), schizophrenia, Huntington's disease, closed head injury, and temporal lobe abnormality (Bigler et al., 1989; Butters et al., 1986; Majdan et al., 1996; Strauss et al., 2006; Tierney et al., 1994) (Mohammadi et al., 2018). Numerous studies have noted that the RAVLT can be effective in identifying patients with dementia of the Alzheimer's type (DAT). Tierney et al. showed that the RAVLT is useful for distinguishing between dementia caused by PD, DAT, and normal aging (Tierney et al., 1994).

In another study, Tierney and coworkers showed that patients with early AD and subcortical ischemic vascular dementia can be accurately distinguished by the recognition memory subtest of the RAVLT and the Controlled Oral Word Association Test (Tierney et al., 2001). A diagnostic tool in AD by Haddad and Nussbaum suggested that patients with a positive RAVLT learning curve are more likely to benefit from rehabilitation compared to patients who have a lower curve (Haddad & Nussbaum, 1990).

In addition, clinical diagnosis of the AD using the RAVLT showed that the delayed recall subtest of the RAVLT and the mental control of the Wechsler memory scale have the same precision and sensitivity to distinguish probable AD. They stated that probable AD can be predicted with high accuracy by these tests (Tierney et al., 1996). Furthermore, the RAVLT has been presented as a useful tool for the differential diagnosis of patients with an increased risk of AD, such as patients with subjective memory problems (Estévez-González et al., 2003).

The pathological diagnosis value of dementia can determine the progression of the disease (Bruscoli & Lovestone, 2004; Mitchell & Shiri-Feshki, 2008, 2009). Early and accurate prediction of the progression of early AD during the course of the disease substantially improves treatment efficiency. In this study, we aim to assess the function of patients with MCI and miAD, and their AVL disabilities by the RAVLT. We also evaluate the possibility of the RAVLT in early diagnosis of miAD.

## 2. Methods

### *2.1. Participants*

A total of 80 subjects, including 20 patients with miAD and 30 with MCI were compared with 30 CNs. The patients were selected from the Baqiyatallah Hospital (Tehran, Iran), and signed a standard informed consent form given by the local ethics committee. Participants with miAD and MCI were referred to the clinic and identified by expert neurologists. miAD patients were diagnosed according to the Diagnostic and Statistical Manual of Mental Disorders (DSM-IV) . The identification of MCI patients was done by an experienced neurologist according to DSM-IV and the research of Petersen and Negash (2008) using the following criteria: (1) existence of any subjective memory complaints; (2) normal orientation and general cognitive performance as described by Folstein′s MMSE score (Folstein et al., 1975); (3) normal score on language ability, evaluated by the Boston Naming Test (BNT); and (4) not knowing the cause of the memory deficit, use of a medication known to affect memory performance, or a major medical or neurological illness. The CN group was essentially determined as individuals with the following criteria: (1) no recognized subjective memory complaints; (2) no approved detectable cognitive and memory deficiencies; (3) MMSE score ≥ 27; (4) normal social operative in the community; and (5) no active neurologic or psychiatric illnesses. The subjects' data was presented in Table 1. The Ethical Committee of the Baqiyatallah University of Medical Sciences approved the study design.

**Table 1. Demographic and neuropsychological characteristics of the groups**

| Characteristics | Normal controls (n=30) | Participants with miAD (n=20) | Participants with MCI (n=30) |
|---|---|---|---|
| Age (years) | 67.033 ± 4.24 | 73.65 ± 2.476[###] | 67.166 ± 3.19[***] |
| MMSE score | 28.06 ± 1.01 | 19.3 ± 1.031[###] | 26.16 ± 1.08[***,###] |
| Education (years) | 11.250 ± 2.51 | 10.26 ± 2.94 | 10.46 ± 2.67 |
| Right/left handedness (n) | 30:0 | 20:0 | 30:0 |
| Women/men (n) | 15:15 | 13 :7 | 19 :11 |
| Duration of disorder (years) | - | 4.05±0.825 | 1.7±0.651[*] |

MCI = Mild Cognitive Impairment; MMSE, total score; Handedness, based on Edinburgh Handedness Inventory; Values are mean ± SD.
[***]$p < 0.001$, statistical comparisons (ANOVA with Tukey HSD post hoc): to the miAD group.
[###]$p < 0.001$, statistical comparisons (ANOVA with Tukey post hoc): to the control group.

### *2.2. Results for the Rey Auditory-Verbal Learning Test (RAVLT)*

The RAVLT was administered immediately following the MMSE and medical history interview for all participants (Rey, 1941; Schmidt, 1996). The RAVLT has two lists (A and B) with 15 words each. List A with 15 words was read loudly by the examiner (with an interval of a second between each word) for five sequential trials (Trials 1 to 5), and each trial was followed by a free-recall test. Participants were told that they would hear a list of 15 words and should listen attentively since they would be requested to retrieve as many of the words as possible. The sequence of the words remained constant across trials. After the fifth trial, a list (List B) of interferer words (15 words) was read out to the subject by the examiner like in previous trials, followed by its free-recall test (Trial 6). Immediately after the sixth trial, the examiner asked the participant to recall the words in List A (Trial 7; immediate recall). After a 30-minute interval, the examiner asked the participant to retrieve the words that belonged to List A once again (Trial 8; delayed recall). After Trial 8, the participant was presented with a test of memory recognition (Trial 9), in which a list containing 15 words from List A, 15 words from List B, and 20 distracting words (phonologically or semantically similar to the words in list A and B) were read out to the participant by the examiner. After each word was read loudly, the participant was requested to mark it and say whether the word was from List A or not. The total time for use of the RAVLT ranges from 35 to 40 minutes. The following indices were computed according to previously described indices: (1) The correct responses were given in each trial; (2) total free recall score: the sum of all correct responses given in the five consecutive trials; (3) verbal learning: difference in the number of words correctly recalled after the first and the fifth readings; (4) verbal forgetting: difference in the number of words correctly recalled after the fifth reading and at the delayed recall test (Trial 8); and (5) learning curve: learning effects from the first trial to the last trail (Trials 1 to 9) (Gainotti & Marra, 1994; Incalzi et al., 1995).

## 2.3. Analysis Methods

The analysis was administered using the one-way ANOVA test to measure differences among groups in demographic variables. For multiple comparisons of significant effects, a post hoc Tukey's HSD test was carried out. Pearson's correlation coefficients were applied to measure the relationship between the MMSE and RAVLT variables. All of the analysis was two-tailed, and the significance level was defined as $P < 0.05$. Statistical computations were carried out by the Statistical Package for the Social Sciences, version 22 (SPSS 22).

## 3. Results

A one-way ANOVA analysis of the demographic variables showed significant differences in age, MMSE scores, and duration of disorder among groups ($p < 0.001$). There was no difference in handedness and years of education among groups ($p > 0.05$). The miAD and MCI groups were predominantly female (65% and 63%, respectively), whereas the CN group had an equal number of men and women. The demographic characteristics of the participants are presented in Table 1.

There were significant differences among the groups in their performance in five trials, total scores, Trial 6 (List B; interference), Trial 7 (immediate recall), Trial 8 (delayed recall), memory recognition (Trial 9), verbal learning (difference between Trials 5 and 1), and verbal forgetting (difference between Trials 8 and 5) ($p < 0.001$). A post hoc Tukey's HSD test revealed that the miAD group had lower scores compared to the CN and MCI groups in all the RAVLT subtests ($p < 0.001$). Similarly, the MCI group had lower scores in these variables than the miAD and CN groups ($p < 0.001$). Analysis of verbal learning showed that there was a significant difference between CN and other groups ($p < 0.001$), but there was no significant difference between the MCI and miAD subjects ($p > 0.05$). Likewise, the verbal forgetting analysis showed that there were no differences among the groups ($p > 0.05$) (Table 2).

Table 2. Performance of all groups on the RAVLT

| Characteristics | Normal controls (n=30) | Participants with miAD (n=20) | Participants with MCI (n=30) |
|---|---|---|---|
| trial 1 (Max 15 words) | 5.9 ± 1.21 | 0.85 ± 0.587### | 2.6 ± 0.96###,*** |
| trial 2 (Max 15 words) | 7.53 ± 1.67 | 1.2 ± 0.615### | 3.6 ± 13###,*** |
| trial 3 (Max 15 words) | 8.6 ± 1.45 | 1.8 ± 0.615### | 4.13 ± 1.54###,*** |
| trial 4 (Max 15 words) | 9.9 ± 1.56 | 2.4 ± 0.502### | 4.8 ± 1.06###,*** |
| trial 5 (Max 15 words) | 10.53 ± 1.75 | 2.55 ± 0.51### | 4.73 ± 1.41###,*** |
| List B: Interference trial 6 (Max 15 words) | 3.88 ± 1.26 | 0.7 ± 0.732### | 1.93 ± 0.9###,*** |
| *Recall score:* Total trials 1 to 5 (Max 60 words) | 42.46 ± 5.87 | 8.8 ± 1.673### | 19.86 ± 5.25###,*** |
| *Immediate recall score:* trial 7 (Max 15 words) | 8.2 ± 2.18 | 0.9 ± 0.718### | 3.23 ± 1.04###,*** |
| *Delayed recall score:* trial 8 after 30 minutes delay (Max 15 words) | 7.666 ± 2.122 | 0.4 ± 0.598### | 1.4 ± 0.96###,*** |
| *Verbal forgetting:* Difference between trials 8 and 5 | 2.66 ± 1.24 | 2.15 ± 0.812 | 3.33 ± 0.8 |
| *Verbal learning:* Difference between trials 5 and 1 | 4.63 ± 1.9 | 1.75 ± 0.85### | 2.1 ± 1.02### |
| Memory recognition (trial 9) | 12.9 ± 1.42 | 1.85 ± 1.136### | 7.43 ± 1.52###,*** |

Values are mean±SD.
***p < 0.001, statistical comparisons (ANOVA with Tukey HSD post hoc): to the miAD group.
###p < 0.001, statistical comparisons (ANOVA with Tukey post hoc): to the control group.

### 3.1. Correlation between MMSE scores and subtests of the RAVLT

The correlation between the scores of MMSE and performance on total recall, interference, immediate recall, delayed recall, verbal forgetting, verbal learning, and memory recognition was measured in each group using Pearson's correlation coefficient. In healthy cognitive controls, the correlation analysis showed that there existed a significant and direct relation among MMSE score and total recall (r = 0.457, p = 0.011), immediate recall (r = 0.506, p = 0.004), delayed recall (r = 0.523, p = 0.003), and verbal learning scores (r = 0.477, p = 0.008), but there were no significant correlations among MMSE score and interference (r = 0.305, p = 0.101), verbal forgetting (r = -0.150, p = 0.429), and memory recognition scores (r = 0.100, p = 0.598) (Table 3). The correlation analysis in the MCI group demonstrated that there were no significant relations among MMSE score and total recall (r = 0.113, p = 0.553), interference (r = 0.152, p = 0.423), immediate recall (r = 0.056, p = 0.769), delayed recall (r = 0.057, p = 0.766), verbal forgetting (r = 0.159, p = 0.400), verbal learning (r = 0.077, p = 0.685), and memory recognition scores (r = 0.080, p = 0.675) (Table 4). Also, correlation analysis in the miAD group showed that there were no significant relations between MMSE score and other variations, except memory recognition (r = -0.454, p = 0.045) (Table 5).

**Table 3. Pearson Correlation coefficients in NC subjects**

| Test | MMSE | P Value |
|---|---|---|
| Interference | 0.305 | 0.101 |
| *Recall score* | 0.457* | 0.011 |
| *Immediate recall score* | 0.506** | 0.004 |
| *Delayed recall score* | 0.523** | 0.003 |
| *Verbal forgetting* | -0.150 | 0.429 |
| *Verbal learning* | 0.477** | 0.008 |
| Memory recognition | 0.100 | 0.598 |

**Correlation is significant at the 0.01 level (2-tailed);
*Correlation is significant at the 0.05 level (2-tailed).

**Table 4. Pearson Correlation coefficients in MCI subjects**

| Test | MMSE | P Value |
|---|---|---|
| Interference | 0.152 | 0.423 |
| *Recall score* | 0.113 | 0.553 |
| *Immediate recall score* | 0.056 | 0.769 |
| *Delayed recall score* | 0.057 | 0.766 |
| *Verbal forgetting* | 0.159 | 0.400 |
| *Verbal learning* | 0.077 | 0.685 |
| Memory recognition | 0.080 | 0.675 |

Correlation was not significant.

**Table 5. Pearson Correlation coefficients in miAD subjects**

| Test | MMSE | P Value |
|---|---|---|
| Interference | -0.153 | 0.519 |
| *Recall score* | -0.024 | 0.919 |
| *Immediate recall score* | -0.171 | 0.472 |
| *Delayed recall score* | -0.119 | 0.616 |
| *Verbal forgetting* | -0.03 | 0.9 |
| *Verbal learning* | 0.006 | 0.979 |
| Memory recognition | 0.454* | 0.045 |

Correlation is significant at the 0.05 level (2-tailed).

## 4. Discussion

This study evaluated the performance of miAD and MCI patients on the RAVLT, and compared their performances with normal cognitive controls. The AVL is commonly measured by the RAVLT in AD and MCI patients, and their subtests were presented as a reliable tool for differentiating the older individuals from AD patients. It has also been revealed that the delayed recall subtest can distinguish the prodromal phase of AD and MCI from each other (Estévez-González et al., 2003; Mohammadi et al., 2018; Tierney et al., 1996). Likewise, the ability of delayed recall to accurately predict probable AD has been previously reported (Tierney et al., 1996). Patients with miAD and MCI performed significantly badly compared to normal cognitive controls in Trials 1−5, interference (Trial 6), immediate recall (Trial 7), delayed recall (Trial 8), memory recognition (Trial 9), and verbal learning but not in verbal forgetting. Consistent with previous research, age and education were related to the RAVLT performance. The average of the RAVLT scores for the patients with MCI was less than that of their age-matched healthy cognitive controls. The results of healthy cognitive control performance on the RAVLT showed that total recall, immediate recall, delayed recall, and verbal learning scores positively correlated with the MMSE score. However, no significant correlation was found among total recall, interference, immediate recall, delayed recall, verbal forgetting, verbal learning scores, and the MMSE score in the patients group. Totally, these findings confirmed that several items of the RAVLT contributed to a differential diagnosis in the prodromal phase of AD.

Antonelli-Incalzi and colleagues showed that the RAVLT indices can be used to differentiate between the very elderly people and AD patients, and that the rates of forgetting, immediate, and delayed RAVLT indices were the powerful discriminant indices (Incalzi et al., 1995). DAT patients exhibited significant impairment compared to closed head injury subjects (Bigler et al., 1989), and patients influenced by depressive ″pseudo dementia″ (Gainotti & Marra, 1994). Estévez-González et al. (2003) concluded that the RAVLT can assist in distinguishing patients with subjective memory complaints who would develop DAT during a few years, and furthermore differentiate among the prodromal phase of AD, MCI, and healthy cognitive aging. Moreover, they declared that the score of zero at the delayed recall trial or percentage of forgetting ≥ 75% in individuals with subjective memory complaint are indicative of plausible DAT in the future. The results of Boone et al. (2005) confirmed that incorporated indices of recognition memory from the RAVLT are effective in recognizing non-credible memory performance in ″real world″ samples and are slightly higher than the 67.2% sensitivity obtained with the standard recognition score. Zeidman and coworkers suggested that a huge number of false alarm errors responded in recognition memory in a frame of an extensive neuropsychological assessment (Zeidman et al., 2008). This may be useful in early diagnosis of MCI (Zeidman & Schweiger et al., 2008). In our study, patients with MCI were significantly lower than healthy cognitive subjects on all components of the RAVLT. Our findings support previous studies which demonstrated that the RAVLT delayed recall can diagnose individuals with DAT, patients with other types of dementia, and control subjects (Mitrushina et al., 1991; Tierney et al., 1996). A logistic regression analysis accurately arranged 94% of the normal controls and demented patients using delayed recall test (Tierney et al., 1996). Moreover, it has been revealed that the RAVLT delayed recall can help predict probable AD with a high degree of accuracy in memory-impaired non-demented patients who were followed longitudinally for two years (Tierney et al., 1996). Estevez-Gonzalez et al. showed that the delayed recall contributes to the early detection of DAT, MCI, and subjects with no evidence of cognitive impairment (Estévez-González et al., 2003). Our study revealed that delayed recall also contributes to early detection of DAT and to the difference between MCI and cognitively normal subjects.

The performance on the RAVLT is crucial for detecting early deterioration in DAT. It has been revealed that the atrophy of medial temporal (hippocampal-parahippocampal formation) is mostly seen in subjects with DAT or MCI, and both the hippocampal and entorhinal cortex volumes were reduced increasingly per annum (Du et al., 2001; Jack et al., 2000; Wolf et al., 2001). Since hippocampal degeneration happens earlier than the start of obvious dementia, atrophy of this area may anticipate the development of future DAT (Convit et al., 2000; Du et al., 2001). Presumably, the left temporal area is essential for the successful performance of the RAVLT, as Kopelman and colleagues demonstrated that the left medial temporal lobe and left prefrontal activation were associated with incremental learning through sequential trials with word stimuli derived from the RAVLT (Kopelman et al., 1998). Also, the atrophic changes of the left medial temporal area would be a proper marker for the early phase of DAT (Wolf et al., 2001; Yamaguchi et al., 2002). Therefore, medial temporal atrophy can describe why RAVLT trials are highly sensitive to a cognitive deterioration in AD.

Finally, our findings confirm that the RAVLT can play a part in the prediction of probable miAD and suggest the coverage of this test in the early evaluation of individuals with subjective memory complaints. According to our outcomes, the most reliable items on the RAVLT were the total recall score (sum of the five learning trials), immediate recall (instantly after interference trial), delayed recall (after 30 minutes delay), verbal learning from Trials 1 to 5, and profile of the learning curve. These measures help distinguish probable miAD, MCI, and normal aging or aging individuals with no evidence of anticipated cognitive impairment.

There are limitations in this study which should be taken into consideration. One of the limitations is the lack of young healthy cognitive controls to compare their performance on the RAVLT with normal aging controls, miAD, and MCI patients. Another limitation is the lack of imaging analysis to explore brain damage in both the groups and compare them with the results of the RAVLT. Despite these limitations, our findings provide beneficial information for clinicians applying the RAVLT. Our results demonstrate that word list learning was impaired in miAD and MCI patients compared to age-matched healthy cognitive controls. However, just poor performance on the RAVLT is not adequate for diagnosing the presence of neurological dysfunction. In fact, interpretation should be done after assessing the general neuropsychological condition along with medical, psychiatric, and historical data.

**Author contributions**

All authors designed the study, supervised the data collection and wrote the paper.

**Disclosure statement**

There is no actual or potential conflict of interest regarding this article.

**Acknowledgements**

We express our thankfulness to the subjects who participated in this study. All authors had complete access to all of the data in the study and take responsibility for the integrity of the data and the truth of the data analysis. This research was supported by grant BMSU/960826 from the Neuroscience Research Center, Baqiyatallah University of Medical Sciences.